\documentclass[preprintnumbers,twocolumn,superscriptaddress,
floatfix,showpacs,prd,aps,amsmath,amssymb,nofootinbib]{revtex4}
\usepackage{graphicx}

\newcommand{\ave}[1]{\left\langle #1 \right\rangle}
\usepackage{hyperref}

\begin{document}

\title{Baryon number conservation and the cumulants of the net proton distribution}

\author{A. Bzdak}
\email[E-Mail:]{ABzdak@bnl.gov}
\affiliation{RIKEN BNL Research Center, Brookhaven National Laboratory, Upton, NY 11973, USA}

\author{V. Koch}
\email[E-Mail:]{VKoch@lbl.gov}
\affiliation{Lawrence Berkeley National Laboratory, Berkeley, CA 94720, USA}

\author{V. Skokov}
\email[E-Mail:]{VSkokov@bnl.gov}
\affiliation{Physics Department, Brookhaven National Laboratory, Upton, NY 11973, USA}

\date{\today}

\begin{abstract}
We discuss the modification of the cumulants of the net baryon and net proton
distributions due to the global conservation of baryon number in heavy-ion
collisions. Corresponding probability distributions and their cumulants are
derived analytically.
We show that the conservation of baryon number results in a substantial decrease
of higher order cumulants. Based on our studies, we propose an
observable that is insensitive to the modifications due to  baryon
number conservation.
\end{abstract}

\maketitle

\section{Introduction}

The phase structure of the strong interactions, Quantum Chromodynamics (QCD),
has been studied theoretically and experimentally for many years.

On the theory side, the problem has been investigated within various models
and addressed systematically in first-principles lattice QCD (LQCD) calculations. 
Due to the sign problem, rigorous results from LQCD on the phase
structure are presently available only at vanishing net baryon density, where
it has been established that QCD exhibits an analytic crossover transition %
\cite{Aoki:2006we} with a pseudo-critical temperature of $T_c\simeq 160 \,%
\mathrm{MeV}$ \cite{Aoki:2006br,Bazavov:2011nk}.

Exploratory LQCD studies at finite net baryon density~\cite{deForcrand:2002ci,D'Elia:2002gd,Allton:2002zi,Fodor:2004nz}, 
as well as model
results (see, e.g. Ref.~\cite{Asakawa:1989bq}), provide some indications for the existence of a critical end point
(CEP) of a first-order phase transition at finite temperature and
density. 

In the laboratory, the properties of hot and dense QCD matter have been studied
in heavy-ion collisions. Since the location of the CEP and the first-order
coexistence
region is not really known from theory, a search of this phase structure
requires the exploration of the entire experimentally accessible phase
diagram. This is achieved by varying the beam energy and monitoring
observables for possible non-monotonic behavior. A first set of experiments of this
nature has been carried out by the NA49 collaboration \cite%
{Alt:2008ca,:2008vb} at the CERN SPS and is now pursued at somewhat higher
energy in the recent beam energy scan program at RHIC~\cite{Aggarwal:2010wy}. 
Both the CEP and the first-order phase transition are associated with characteristic
fluctuations -- long range ones for the second-order transition at the CEP, and
possible spinodal instabilities in case of a first-order transition 
\cite{Sasaki:2007db,Randrup:2009gp}. The first measurements by the
NA49 collaboration have  concentrated on the variances, or second-order cumulants, of various
particle ratios and the transverse momentum as was proposed in Ref. \cite%
{Stephanov:1999zu}. Their measurements showed hardly any deviation from the
expected Poisson fluctuations of a Hadron Resonance Gas (HRG). 

Recently it
has been realized that higher order cumulants, especially of the baryon
number, which serves as an order parameter of the QCD phase transition at
finite density, would be more sensitive to the fluctuations associated with
the second order transition, including the CEP \cite{Stephanov:2008qz}. First,
because higher order cumulants scale with higher powers of the correlation
length, a finite and limited increase of the correlation length as a
result of critical slowing down may still be visible. Second, since the
baryon number is a conserved quantity its fluctuations are less modified by
the final state interaction in the hadronic phase
\cite{Koch:2008ia}. Furthermore, model calculations 
found that the kurtosis, i.e., the ratio of
the fourth- over the second-order cumulant, is negative at high baryon chemical
potential, close to the chiral crossover line above the CEP~\cite%
{Skokov:2010uh}. In Ref.~\cite{Stephanov:2011pb}, the lines, where the
kurtosis changes sign, were obtained from universality arguments in the
critical region of the CEP. These theoretical predictions suggest a
non-monotonic behavior of the kurtosis of the baryon number or electric charge distributions 
as a function of the collision energy if chemical freeze-out happens close to the CEP.

In addition, higher order cumulants also provide sensitive information about
the crossover  at vanishing baryon density. As shown in Ref.~\cite%
{Friman:2011pf}, the sixth-order cumulant of the net baryon distribution is
negative close to the crossover line. Preliminary LQCD results~\cite%
{Schmidt:2010xm} also indicate negative values of the sixth-order cumulant
close to the crossover temperature at zero baryon chemical potential. Thus,
by measuring the sixth-order cumulants, one may relate the crossover and the
freeze-out line at non zero chemical potential and provide information
on an approximate position of the possible CEP.

When comparing theoretical (model or LQCD) predictions to the experimental
data, one has to keep in mind that the singular behavior of fluctuations is
predicted in the grand canonical formulation of thermodynamics where 
conservation laws are imposed only on the average. Consequently, to address the
same physics experimentally, one is required to approximately achieve
conditions of the grand canonical ensemble, i.e., to study fluctuations in a
restricted phase space \footnote{Clearly, if {\em all} particles are
  observed, the baryon number will not fluctuate.}. 
This might be done by performing appropriate cuts in
the rapidity and/or transverse momentum of detected
particles. Clearly, the smaller the fraction of observed particles the smaller 
is the effect of global baryon/charge conservation. This was also demonstrated in Ref.~\cite{Schuster:2009jv} using the UrQMD model. However, care has to be taken 
that these cuts do not destroy the underlying correlations responsible for 
the physics one tries to access. Therefore, a fine balance between the 
need to suppress the effects of conservation laws and the requirement 
to preserve the dynamical correlations has to be found~\cite{Koch:2008ia}. 
The subsequent studies, we believe, will help in achieving this nontrivial task. 

In this paper, we will explore to what extent global baryon
  conservation modifies cumulants of net baryon and net proton
  distributions as well as their ratios. To this end we consider a
  system where the only correlations are due to global baryon number
  conservation. Therefore, we start with Poisson statistics for
  baryons and anti-baryons and, subsequently, enforce baryon number
  conservation. Since the  Hadron Resonance Gas
(HRG) model in the classical or Boltzmann approximation is also
governed by Poisson statistics for the baryon and anti-baryons, our
results may be directly applied to the HRG, which is commonly used
as a theoretical baseline for the analysis of heavy-ion collisions \footnote{For baryons 
and anti-baryons the classical (Boltzmann) approximation is justified due to a large ratio 
of mass to temperature.}.  
In other words, our approach is equivalent to a treatment of the HRG
in the so called canonical ensemble~\cite{Cleymans:1990mn} with respect to the
baryon number. A canonical treatment of the HRG with respect to
strangeness has been reported in the literature, e.g., in
Refs.~\cite{Cleymans:1990mn,Koch:2005pk}. The effect of a globally conserved
charge on the variance of the charge distribution has been studied in
the same framework in Refs.~\cite{Bleicher:2000ek,Begun:2004gs}. Here we
will extend these studies to higher order cumulants, where the effects
due to global baryon number conservation are expected to be stronger. 
Therefore, our results  may be considered as an improved
Hadron Resonance Gas prediction for the baryon number cumulants, and
thus provide a baseline with which measurements should be compared in
order to see whether there are {\em additional} dynamical
correlations. Clearly, this baseline can be improved by further
imposing electric charge conservation. However, the number of charged
particles in high energy heavy-ion collision is considerably larger
than the number of baryons plus anti-baryons. Therefore, the
corrections due to electric charge conservation will be sub-leading and
only become relevant as the collision energy is reduced. This will be
discussed in detail in Ref.~\cite{paper_2}. 

Ultimately it would be desirable to incorporate the effects of global
baryon number conservation into the various models or preferably into
Lattice QCD. This is a very difficult task, and, therefore, we believe
for the time being the present study will be helpful for the
interpretation of the experimental data.

In the next Section, we derive an analytical formula for the net baryon (and
net proton) probability distribution under constraints imposed by global
baryon number conservation. Also the cumulant generating function, which can be
used to compute cumulants of any order, is derived. In Section III,
we consider properties of the cumulants up to the sixth order and propose a
new observable, which is insensitive to the global conservation of baryon number.
Comments and conclusions are presented in Section IV and Section V,
respectively.

\section{Global baryon conservation}

Before we derive the relevant formulae let us remind ourselves
  what the problem at hand is. On the one hand we need to impose
  baryon number conservation on a system of baryons and anti-baryons
  following an underlying Poisson distribution. On the other hand we
  have to model the finite acceptance in an experiment, since for full
  acceptance the baryon number does not fluctuate. Here, we model the
  finite acceptances simply by a binomial distribution, noting that in
  practice this may be more involved \footnote{The use of a binomial
    distribution is correct if there are no correlations among 
    baryons and anti-baryons other than the global conservation of
    baryon number which we take into account explicitly. See section
    \ref{sec:comments} for further discussion on this point.}. 

  These two tasks may be done in any
  order, i.e. one first derives the distribution for all particles subject
  to baryon number conservation and then folds with the binomial
  distribution or vice verse. Here we chose to first separate the
  system into observed and unobserved particles based on the binomial
  distribution. This is straightforward since folding a Poisson
  distribution with a binomial results again in a Poisson
  distribution. Next we impose baryon number conservation on {\em all}
  particles, observed and unobserved. 

To get started, let us remind ourselves that the probability distribution of the difference
of two independent random variables, each drawn from a Poisson
distribution, is the so called Skellam distribution. Therefore, in our
approach as well as in the HRG in the Boltzmann limit, in the absence
of baryon number conservation the net baryon number is distributed
according to the Skellam distribution (see e.g.~\cite{BraunMunzinger:2011dn,BraunMunzinger:2011ta}) . Thus, in the following we
will generalize the Skellam probability distribution by
imposing global baryon number conservation. For the sake of simplicity, we
perform our derivation for the net baryon number, and later we generalize
our result to net protons.

Suppose we have on average $\ave{N_B}$ baryons and
$\ave{N_{\bar{B}}}$ anti-baryons in the full
phase space and an average net baryon number of
$B=\ave{N_B}-\ave{N_{\bar{B}}}$. According to our assumptions, both $N_{B}$ and
$N_{\bar{B}}$ follow a Poisson distribution. In order to model the
finite acceptance we next split the full phase space into two 
subsystems, one representing the measured particles,
and one representing the unobserved rest of the particles. 
In the absence of any correlations particles are
distributed between two subsystems according to a binomial
distribution, where the probability $p_B$ ($p_{\bar{B}}$) to observe
a baryon (antibaryon) is simply given by the fraction of the average 
number of observed baryons (antibaryons) to the average number of 
baryons (antibaryons) in the full 
phase space~\footnote{Obviously $0\leq p_{B,\bar{B}}\leq 1$.}. 
Consequently in both subsystems baryons and antibaryons are distributed according to Poisson
distributions with appropriate means. 

As a result, the probability to observe $n_{1}$
net baryons in the measured phase space is again given by the Skellam distribution~\cite{BraunMunzinger:2011dn,BraunMunzinger:2011ta} 
\begin{equation}
P_{1}(n_{1})=\mathcal{N}_{1}\left( \frac{p_{B}\left\langle
N_{B}\right\rangle }{p_{\bar{B}}\left\langle N_{\bar{B}}\right\rangle }%
\right) ^{n_{1}/2}I_{n_{1}}\left( 2z\sqrt{p_{B}p_{\bar{B}}}\right) ,
\label{P1}
\end{equation}%
and analogously in the unmeasured phase space, we have 
\begin{eqnarray}
P_{2}(n_{2}) &=&\mathcal{N}_{2}\left[ \frac{(1-p_{B})\left\langle
N_{B}\right\rangle }{(1-p_{\bar{B}})\left\langle N_{\bar{B}}\right\rangle }%
\right] ^{n_{2}/2}  \nonumber \\
&&\times I_{n_{2}}\left( 2z\sqrt{(1-p_{B})(1-p_{\bar{B}})}\right) ,
\label{P2}
\end{eqnarray}%
where $\mathcal{N}_{1,2}$ are unimportant normalization constants and%
\begin{equation}
z=\sqrt{\left\langle N_{B}\right\rangle \left\langle N_{\bar{B}%
}\right\rangle }.  \label{z}
\end{equation}%
The joint probability to have $n_{1}$ net baryons in the observed subsystem and $n_{2}$
in the unobserved subsystem is given by 
\begin{equation}
P(n_{1},n_{2})=P_{1}(n_{1})P_{2}(n_{2}).  \label{Pjoint}
\end{equation}%
To impose the conservation of baryon number we multiply $P(n_{1},n_{2})$ by $%
\delta _{n_{1}+n_{2},B}$ and sum over all values of the unobserved net
baryon number, $n_{2}$, 
\begin{equation}
P_{B}(n_{1})=\mathcal{N}\sum\nolimits_{n_{2}}P_{1}(n_{1})P_{2}(n_{2})\delta
_{n_{1}+n_{2},B}.  \label{Pn1}
\end{equation}%
The normalization factor $\mathcal{N}$ is fixed from the condition 
\begin{equation}
\sum\nolimits_{n_{1}}P_{B}(n_{1})=1.  \label{Norm}
\end{equation}%
Using Graf's addition formula \footnote{%
Graf's addition formula \cite{Graf} is given by\newline
$\sum\limits_{k}t^{k}I_{k}(x)I_{n-k}(y)=\left( t\frac{y+tx}{x+ty}\right) ^{%
\frac{n}{2}}I_{n}\left( \sqrt{x^{2}+y^{2}+\frac{1+t^{2}}{t}xy}\right) $,
which for $t=1$ reduces to $\sum\limits_{k}I_{k}(x)I_{n-k}(y)=I_{n}(x+y)$.}
we obtain the net baryon probability distribution 
\begin{eqnarray}
P_{B}(n) &=&\left( \frac{p_{B}}{p_{\bar{B}}}\right) ^{n/2}\left( \frac{%
1-p_{B}}{1-p_{\bar{B}}}\right) ^{(B-n)/2}  \label{PB1finalG} \\
&&\times \frac{I_{n}\left( 2z\sqrt{p_{B}p_{\bar{B}}}\right) I_{B-n}\left( 2z%
\sqrt{(1-p_{B})(1-p_{\bar{B}})}\right) }{I_{B}(2z)},  \nonumber
\end{eqnarray}%
where $z$ is given in Eq. (\ref{z}). A detailed derivation of this result
will be shown elsewhere \cite{paper_2}. 
The cumulant
generating function $g(t)$ for the cumulants $c_k$ 
\begin{equation}
{g}(t)=\sum_{k=1}^{\infty }c_{k}\frac{t^{k}}{k!}  \label{g-sum}
\end{equation}
is given by 
\begin{eqnarray}
{g}(t) &=&\ln \left( \sum\nolimits_{n}P_{B}(n)e^{nt}\right)   \nonumber \\
&=&\ln \left[ \left( \frac{{q}_{+}}{{q}_{-}}\right) ^{B/2}\frac{I_{B}(2z%
\sqrt{{q}_{+}{q}_{-}})}{I_{B}(2z)}\right] ,  \label{CGFG}
\end{eqnarray}
where ${q}_{+}=1-p_{B}+p_{B}e^{t}$ and ${q}_{-}=1-p_{\bar{B}}+p_{\bar{B}%
}e^{-t}$. 

Finally, the probability distribution for net protons results from the
observation that all baryons other than protons (e.g. neutrons) may be
considered as unobserved baryons. Thus, to obtain analogues of
Eqs. (\ref{PB1finalG}) and (\ref{CGFG}) for net protons, one simply defines the binomial probability as follows
\begin{equation}
p_{B}=\frac{\langle n_{B}\rangle }{\langle N_{B}\rangle }\rightarrow \frac{%
\langle n_{p}\rangle }{\langle N_{B}\rangle },
\end{equation}%
where $\langle n_{p}\rangle $ is the mean number of observed protons (and analogously for $p_{\bar{B}})$ \footnote{Since there are at least as many neutrons 
in the system, $\langle n_{p}\rangle / \langle N_{B}\rangle \leq \frac{1}{2}$.}.

\begin{figure}[t]
\includegraphics[width=7.5cm]{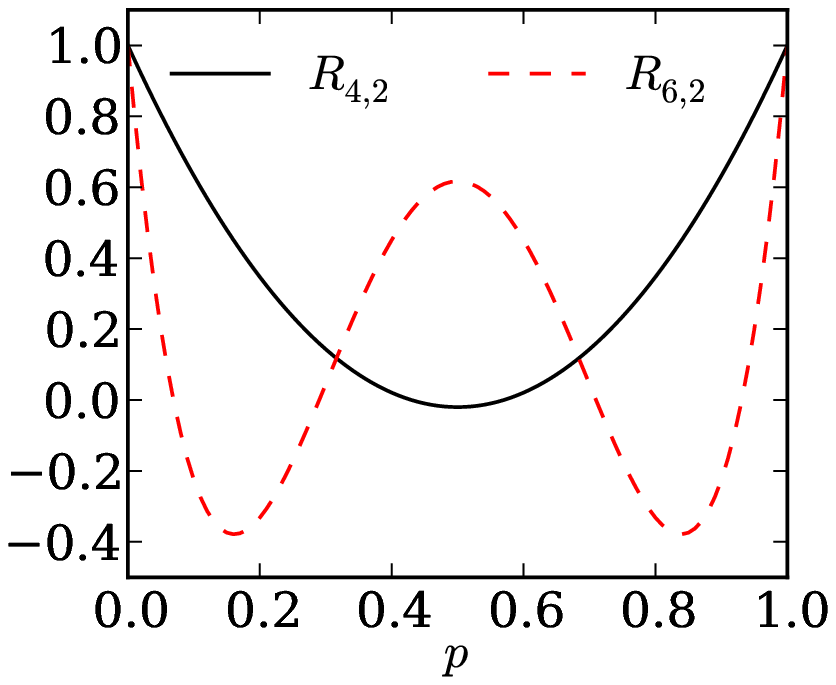} %\hspace{2.5cm}
\includegraphics[width=7.5cm]{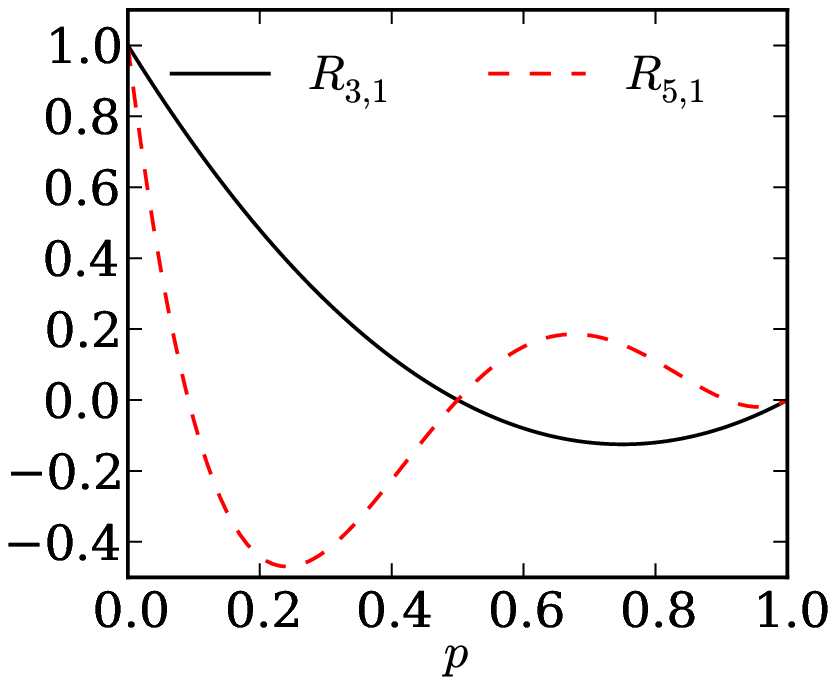}
\caption{Ratios of odd and even order cumulants as a function of the
fraction of measured baryons, $p$. The parameters are $B=300$, $\left\langle
N_{B}\right\rangle =400$ and $\left\langle N_{\bar{B}}\right\rangle =100$.}
\label{Feven}
\end{figure}

\section{Cumulants}

In this section we present the cumulants in the case where $p_{B}=p_{\bar{B}}=p$.
They result from Eq.~(\ref{CGFG}) by taking the appropriate number of derivatives
with respect to $t$ and setting $t$ to $0$.

As seen from Eq.~(\ref{CGFG}), derivatives will generate Bessel
functions of various order. Those can be simplified by using the known properties of the Bessel functions \footnote{$I_{k}(x)=\frac{x}{%
2k}\left[ I_{k-1}(x)-I_{k+1}(x)\right] $.} as well as the 
expression for the average number of baryons and anti-baryons in the
case of baryon number conservation, $\ave{N_B}_C$ and
$\ave{N_{\bar{B}}}_C$, respectively, which will be discussed in the next section (see
Eq.~(\ref{Mean})). 
For the first three even
order cumulants we  obtain%
\begin{eqnarray}
c_{2} &=&p(1-p)\ave{N}_C,  \label{c2} \\
c_{4} &=&c_{2}+3(p^{2}q^{2}B^{2}-c_{2}^{2})+6pq(2z^{2}pq-c_{2}), \\
c_{6} &=&c_{4}+4(c_{4}-c_{2})-10(2pq+c_{2})(c_{4}-c_{2})  \nonumber \\
&&-30pq(p^{2}q^{2}B^{2}+c_{2}^{2}),  \label{c6}
\end{eqnarray}
where $\ave{N}_C=\langle N_{B}\rangle_C +\langle N_{\bar{B}}\rangle_C $, $q=1-p$ and $z$
is defined in Eq.~(\ref{z}).
Note that in the limit $p\rightarrow 0$
one obtains $c_{6}\approx c_{4}\approx c_{2}\approx p\ave{N}_C$. It is worth 
mentioning that in case of the Skellam distribution, i.e., disregarding the conservation law,
we have $c_{6}^{S}=c_{4}^{S}=c_{2}^{S}=p\ave{N}$.

The odd order cumulants can be expressed by polynomials in $p$%
\begin{eqnarray}
c_{1} &=&pB,  \label{odd_c1} \\
c_{3} &=&c_{1}(1-p)(1-2p), \\
c_{5} &=&c_{3}\left( 1-12p(1-p)\right) .  \label{odd_c5}
\end{eqnarray}%
For the Skellam distribution, $c_{5}^{S}=c_{3}^{S}=c_{1}^{S}=pB$. 
As seen
from Eqs. (\ref{odd_c1}-\ref{odd_c5}) the odd order cumulants are linear in $%
B$ and do not depend on $z$. Thus, their ratios  are uniquely defined in
terms of $p$. 

\begin{figure}[b]
\includegraphics[width=7.7cm]{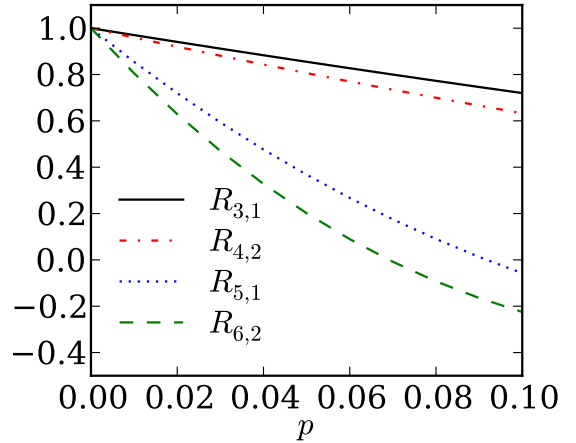}
\caption{Ratios of odd and even order cumulants as a function of the
fraction of measured baryons, $p$ in the range of values, which are of
experiment interest. The parameters are $B=300$, $\left\langle
N_{B}\right\rangle =400$ and $\left\langle N_{\bar{B}}\right\rangle =100$.}
\label{Fzoom}
\end{figure}

Next let us define the ratio $R_{n,m}$ as
\begin{equation}
R_{n,m}=\frac{c_{n}}{c_{m}}.  \label{R}
\end{equation}
In Figs. $1$ and $2$, we show the dependence of the ratios of cumulants, $%
R_{n,m} $ on $p$ for realistic values of $B$ and $\langle N_{B,\bar{B}%
}\rangle $. The ratios of even cumulants are symmetric with respect to $%
p\rightarrow 1-p$ as seen from Eqs. (\ref{c2}-\ref{c6}). We note that
for the Skellam distribution the ratios shown in Figs. 1 and 2 are
unity. Therefore, we observe substantial corrections due to baryon
number conservation.

As already mentioned, the ratios of the
odd order cumulants depend only on $p$. This allows us to construct the
following combination 
\begin{equation}
D=R_{5,1}-R_{3,1}\left[ 1-\frac{3}{4}(1+\gamma )(3-\gamma )\right] ,
\label{D}
\end{equation}
such that $D=0$ for the baryon conservation corrected 
distribution $P_{B}(n)$, Eq. (\ref{PB1finalG}), for any 
values of $p$, $z$ and $B$. Here, $\gamma =\pm \sqrt{1+8R_{3,1}}$. 
The upper (lower) sign should be
taken for $p<3/4$ ($p>3/4$)~\footnote{For an analysis of experimental data, 
the case with $p<3/4$ should be considered.}.
Also, $D=0$ for the Skellam
distribution. Therefore, a deviation of $D$ from zero may indicate
physics, that is not related to global baryon conservation.

\section{Discussion and comments}
\label{sec:comments}

Several comments are in order regarding our results obtained in the previous sections: 

\begin{enumerate}

\item The distribution (\ref{PB1finalG}) depends on $z=\sqrt{\langle
N_{B}\rangle \langle N_{\bar{B}}\rangle }$, where $\langle N_{B}\rangle $ ($%
\langle N_{\bar{B}}\rangle $) is the total baryon (antibaryon) number
present in the Skellam distributions (\ref{P1}) and (\ref{P2}). Thus, $\langle
N_{B}\rangle $ ($\langle N_{\bar{B}}\rangle $) is related to the system
without baryon conservation. It is natural to expect that baryon
conservation will modify $\langle N_{B}\rangle $ ($\langle N_{\bar{B}%
}\rangle $), however, as we argue below this correction is negligible. 
A straightforward calculation gives
\begin{equation}
\left\langle N_{B,\bar{B}}\right\rangle _{\text{\textrm{C}}}=z\frac{I_{B\mp
1}(2z)}{I_{B}(2z)},  \label{Mean}
\end{equation}%
where the upper (lower) sign corresponds to  $\langle N_{B}\rangle
_{\mathrm{C}}$ ($\langle N_{\bar{B}}\rangle _{\mathrm{C}}$), with
$\ave{N_{B}}_{\mathrm{C}}-\ave{N_{\bar{B}}}_{\mathrm{C}}=B$. Here the
subscript $\ave{\cdot}_C$ refers to averages obtained with full
baryon number conservation. Under the constraint 
$\ave{N_B}-\ave{N_{\bar{B}}}=B$, one can express $z$ in terms of $\langle N_{B,\bar{%
B}}\rangle _{\mathrm{C}}$, and to a very good approximation we find
\begin{equation}
z\approx \sqrt{\langle N_{B}\rangle _{\mathrm{C}} \cdot \langle N_{\bar{B}}\rangle
_{\mathrm{C}}}.  \label{zaprox}
\end{equation}%
Using the properties of the modified Bessel
functions, one can show that corrections to Eq.~(\ref{zaprox}) are important
only if both $B$ and $\langle N_{B}\rangle _{\mathrm{C}} \cdot \langle N_{\bar{B}%
}\rangle _{\mathrm{C}}$ simultaneously assume value of the order of one or
smaller. This is never the case in heavy-ion collisions. Relation
(\ref{zaprox}) together with the  requirement that
$\ave{N_B}-\ave{N_{\bar{B}}}=B$  ensures that $\ave{N_B} \approx \ave{N_{B}}_{\mathrm{C}}$ and
$\ave{N_{\bar{B}}} \approx \ave{N_{\bar{B}}}_{\mathrm{C}}$ to very good precision. 
The same identities 
also hold if we only consider protons. Therefore, the formalism
developed in the previous section is of a great phenomenological value since
it allows to calculate the effect of baryon number conservation on
the probability distribution and its cumulants given experimentally
determined average yields. This will be further elaborated in Ref.~\cite{paper_2}.

\item We have shown that the odd order cumulants do not depend on $\langle N_{B,%
\bar{B}}\rangle $, their ratios are independent of $B$ and 
uniquely defined by one parameter, the fraction of observed baryons
(protons), $p$. This turns out to be very useful for the
phenomenological analysis of experimental data. For example, chiral model calculations at non-zero baryon densities show that both $R_{3,1}
$ and $R_{5,1}$ are non trivial functions of temperature and chemical
potential close to the crossover and the CEP. This is demonstrated in Fig.~3, 
where we present the results obtained in the Polyakov loop-extended
Quark-Meson model~\cite{Skokov:2010uh} for $R_{3,1}$ and $R_{5,1}$. We also
show the new observable $D$, see Eq. (\ref{D}), which exhibits strong,
temperature-dependent deviations from the baseline of $D=0$, even 
for
temperatures below the pseudo-critical one, $T<T_{pc}$. Therefore, effects due
to a possible phase transition should be accessible in experiment via an
analysis of this new observable~$D$. 
\begin{figure}[h]
\includegraphics[width=7.5cm]{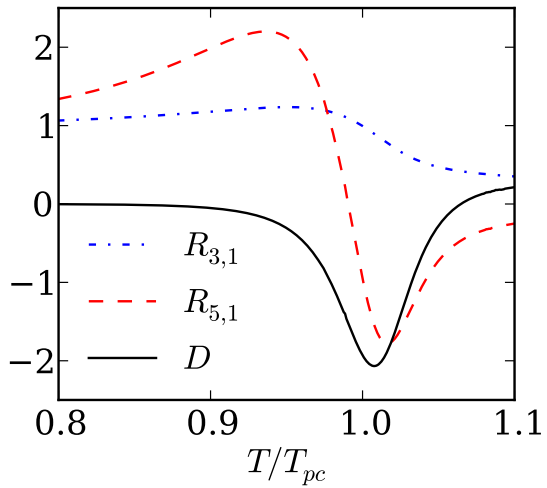}
\caption{The ratios $R_{3,1}$, $R_{5,1}$ and $D$ as a function of
temperature in the PQM model. The calculations are performed along the line of fixed $%
\protect\mu _{B}/T\approx 0.5$. $T_{pc}$ is the crossover 
temperature at the corresponding chemical potential. }
\label{Fpqm}
\end{figure}

\item 
For a given experiment, the parameter $p$ can be roughly
estimated. For example for the STAR experiment~\cite{Aggarwal:2010wy}
the mean number of accepted protons in the most central Au-Au 
collision at $\sqrt{s}=200$ GeV in the measured phase
space is approximately $\langle n_{p}\rangle \approx 7$. The total mean
number of protons can be estimated from data on $dN_{p}/dy$ at zero 
rapidity $dN_{p}/dy=35$ \cite{:2008ez}, and assuming flat rapidity
distribution in the range $-3<y<3$. Therefore, $\langle N_{p}\rangle \approx 35\cdot
6=210$. We are, however,
interested in the total baryon number, $\ave{N_B}$.
Therefore, $\langle N_{p}\rangle$ is to be multiplied by some factor $f$ that takes into
account contribution of neutrons, $\Lambda $ and other long living
resonances. We estimated this factor in the thermal model. At the
temperature $T\approx 166$ MeV corresponding to $\sqrt{s}=200$ GeV we obtain 
$f=2.5$, so that $\ave{N_B}\approx 210\cdot 2.5 = 525$. A similar
number for $\ave{N_B}$ can be obtained using BRAHMS data~\cite{Bearden:2003hx}. 
The fraction of measured protons to the total number of baryons 
is $p\approx 7/525 \approx 0.013$.
For this value of $p$ and $B=350$, we obtain $R_{4,2} \approx 0.95$ 
and $R_{6,2} \approx 0.77$.
There is some uncertainty related to the problem that we should only
include those baryons that play a role in the quasi-equilibrium
physics. This number is, however, difficult to estimate reliably.

\item At sufficiently low energies $\left\langle N_{B}\right\rangle \gg
\left\langle N_{\bar{B}}\right\rangle$ and $z\ll B$,  
the cumulant generating function~(\ref{CGFG}) reduces~to~\footnote{%
For  $2x\ll\sqrt{k+1}$, $I_{k}(2x)\approx x^{k}/\Gamma (k+1)$.} 
\begin{equation}
\tilde{g}(t)
=  B\ln \left[ 1-p(1-e^{t})\right] .  \label{g-limit}
\end{equation}%
In this case all ratios of the cumulants depend only on the fraction of observed
baryons (protons) $p$. Taking $\langle n_{p}\rangle = 9$ and $%
\langle N_{B}\rangle = 350$ (number of participants in central
collisions) we obtain $p\approx 0.026$ and in consequence $%
R_{4,2}\approx 0.85$ and $R_{6,2}\approx 0.32$.

\item In this paper we have disregarded other conservation laws, e.g. electric
charge conservation, which is expected to play a significant role at low
energies. We believe, that at a collision energy higher than $10$ GeV,
energy, momentum and electric charge conservation can be neglected owing to
high abundance of pions. A detailed investigation of electric charge
conservation will be reported elsewhere \cite{paper_2}.

\item We also have neglected non-equilibrium effects, effects of interactions,
volume fluctuations \cite{Skokov:2012ds} resulting, e.g., from
centrality fluctuations, etc. 
For example, here we have modeled the acceptance of baryons and
  anti-baryons simply by a binomial distribution. This is only correct
  if there are no correlations among baryons and anti-baryon other than
  the global baryon number conservation, which we have accounted for explicitly
  in our calculations. In reality, one could very well imagine that
  the stopping of the baryons from the colliding nuclei may give rise
  to correlations and thus fluctuations of the baryon number which are
  not accounted for by the binomial distribution used in our
  calculation. These additional fluctuations  will
  contribute to the various cumulants and thus need
  to be understood in order to extract any signal for critical
  fluctuations. Consequently, our proposed observable, $D$, will also deviate
  from zero, since $D$ is designed to remove only correlations due to
  global baryon number conservation. This in turn  may be utilized to study
  baryon and anti-baryon correlations and fluctuations due to
  stopping.

\item In the present paper we assume that baryon number is conserved
globally. However, it is plausible that baryon number is conserved locally
(a few units of rapidity). This effect would reduce the effective total
number of (anti)baryons $\ave{N_{B,\bar{B}}}$ and increase the value
of $p_{B,\bar{B}}$ and, consequently, the corrections due to baryon
number conservation.
\end{enumerate}

\bigskip

\section{Conclusions}

We have studied the effects of global baryon conservation on the cumulants of
net baryon and net proton fluctuations. We showed that the cumulants are substantially
suppressed if  global baryon conservation is taken into account. We also
proposed a new observable that is insensitive to global baryon
conservation but changes rapidly at the critical end point or the crossover. 

\bigskip

\section{Acknowledgments}

We thank F. Karsch and L. McLerran for discussions. A.~B. and V.~S. were
supported by Contract No. DE-AC02-98CH10886 with the U. S. Department of
Energy. V.~K. was supported by the Office of Nuclear Physics in the US
Department of Energy's Office of Science under Contract No.
DE-AC02-05CH11231. A.~B. also acknowledges the grant N N202 125437 of the Polish Ministry of Science and Higher Education (2009-2012).

\end{document}